\documentstyle[preprint,epsf,aps]{revtex}
\begin{document}
\title{Entanglement concentration by ordinary linear optical devices
without post-selection}
\author{Wang Xiang-bin\thanks{email: wang$@$qci.jst.go.jp}, 
Fan Heng\thanks{email: fan$@$qci.jst.go.jp} 
\\
        Imai Quantum Computation and Information project, ERATO, Japan Sci. and Tech. Corp.\\
Daini Hongo White Bldg. 201, 5-28-3, Hongo, Bunkyo, Tokyo 113-0033, Japan}

\maketitle 
\begin{abstract}

 Recently, entanglement concentrations have been  experimently demonstrated by 
post-selection ( T. Yamamoto et al, Nature, 421, 343(2003),  and
 Z. Zhao et al, Phys. Rev. Lett., 90,207901(2003) ), i.e., to each individual
outcome state, one has to
destroy it  to know whether it has been purified.
Here we give proposal for entanglement
concentration without any post selection by using only practically existing
linear optical devices.  In particular, a sophisticated photon detector to distinguish one photon or two photons 
is not required.
\end{abstract}
The resource of maximally  entangled state(EPR state) plays a fundamentally 
important role in testing the quantum laws
related to the non-locality\cite{ein} and in many tasks of quantum information 
processing\cite{qip,chuang} such as the quantum teleportation\cite{bennett,bou}, quantum dense
coding\cite{bennett},
entanglement based quantum key distribution\cite{ekert} and quantum computation\cite{chuang}.
So far, it is generally believed that the  two-photon polarized EPR state is
particularly useful in quantum information processing.

If certain type of non-maximally entangled states
are  shared by distant
two parties Alice and Bob initially, the raw states
  can be distilled into highly entangled states
by local quantum operation and classical communication through the  entanglement concentration scheme\cite{puri}. Although  research
on such issues have been extensively done theoretically, 
the feasible experimental
schemes and experimental demonstrations of the entanglement concentration
are rare. So far, some schemes
by linear optical devices have been proposed\cite{panature,ypra} and some 
experimental demonstrations
have also been reported\cite{yamamoto,ustc,ins}. In all these 
schemes\cite{panature,ypra,yamamoto,zhao,ustc}, 
one has to verify that each of 
the indicator light beams contains $exactly$ 
one photon. However, by our current technology, it's not likely to really 
implement a sophisticated detector\cite{det} in the scheme 
to distinguish one photon or two photon  in a light beam. What we can see from a normal photon detector is that whether it is clicked or not. 
When it is clicked, the measured  light beam could
contain either exactly one photon or 2 photons in the previously
proposed entanglement concentration schemes\cite{panature,ypra,yamamoto}.
In a recent experimental report\cite{yamamoto}, the unwanted events that
one indicator beam contain two photons are excluded by post selection: Both 
the two indicator beams and the two outcome beams are measured. If all the 4 detectors are clicked, the maximally entangled state must have been created
on the two outcome beams. Such a post selection  destroies the
outcome itself.  That is to say, limited by the available technology
of photon detector, one can only verify a maximally 
entangled state by totally destroying that state. This means, without a
sophisticated photon detector, the set-up\cite{yamamoto}
is not supposed to really produce any maximally entangled state through the 
entanglement concentration, even though the requested raw states are supplied
deterministically.  A similar drawback also appears in\cite{panature,zhao,ins}
. 
The very recent experiment\cite{ustc} also relies on
 post-selection.

So far non-post-selection entanglement concentration in polarization
space with linear optical devices has never been proposed,  
though there are some studies on the possibility
 of the entanglement 
concentration to continuous variable states through the Gaussification
scheme\cite{eisert,eisert2}. 

In the situation of Ref.\cite{yamamoto,panature,ustc}, the entanglement is 
in the two
level polarization space and all pairs are equally shared by two remotely
separated parties. This case is rather important because the polarization
entanglement is easy manipulate, e.g., the local rotation operation.
Also the assumption of two  pair states with unknown $identical$ parameters
is reasonable in cases such as that Alice sends two halves of EPR pairs
to Bob
through the same dephasing channel. Note that here each group contains
two identical pairs with unknown parameters, {\it  the parameters for
the pair states in different groups are different}. Now we go to the main 
result of this Letter. 
 
We will show that the following raw state
\begin{eqnarray}\label{key}
|r,\phi\rangle =\frac{1}{1+r^2}(|HH\rangle + r e^{i\phi}|VV\rangle)\otimes
(|HH\rangle + r e^{i\phi}|VV\rangle)
\end{eqnarray} 
can be probabilistically distilled into  maximally entangled state without
post selection, even though only normal photon detectors are used. 
Note that in Eq.(\ref{key}) $r>0$, both $r$ and $\phi$ are unknown 
parameters. In particular, the special case $r=1$ is just  the one
treated  in the recent experiment\cite{yamamoto}.
Since the raw state contains unknown parameters,
 our purification(concentration) scheme is actually to distill the maximum
entanglement from the un-normalized mixed state 
\begin{eqnarray}\label{rq}
\int_0^{\infty}\int_0^\pi  |r,\phi\rangle\langle r,\phi| d r d\phi.
\end{eqnarray} 
However, for simplicity we shall use the pure state 
notation $|r,\phi\rangle$, keeping in the mind that both parameters are totally
unknown.
Consider the  schematic diagram, figure 1. 
This diagram is a modified scheme from the one given by Pan et al in 
Ref.\cite{panature}. However, as we shall see, the modification leads to 
a totally different
result: Our scheme  uniquely produces the even-ready maximally entangled state through 
entanglement purification,
 provided that the requested state in Eq.(\ref{rq}) is supplied.

Our scheme requires the two fold 
coincidence event as the indication that the maximally entangled state
has been
produced on the outcome beams 2',3', i.e. 
whenever both photon detectors $D_x,D_w$click, the
two outcome beams, beam 2' and beam 3' must be in the maximally
entangled state:
\begin{eqnarray}
|\Phi^+\rangle_{2',3'}=\frac{1}{\sqrt 2}(|H\rangle_{2'}|H\rangle_{3'}
|+V\rangle_{2'}|V\rangle_{3'}).
\end{eqnarray}   
Now we show the mathematical details for the claim above.
The polarizing beam splitters transmit the horizontally polarized photons
 and reflect
the vertically polarized photons. For clarity, we use the Schrodinger picture. And we 
assume the non-trivial time evolutions to the light beams only takes 
place in passing through the optical devices. 

Consider Fig.(1).  Suppose initially two remote parties Alice and Bob share two pairs of non-maximally entangled photons as defined by Eq.(\ref{key}), denoted by photon pair 1,2 and photon 3,4 respectively. 
The half wave plate HWP1 here is to change the polarization between 
the horizontal and 
the vertical. After photon
3 and 4 each pass through  HWP1, the state is evolved to:
\begin{eqnarray}
\frac{1}{1+r^2}(|HH\rangle_{12} + r e^{i\phi}|VV\rangle_{12})\otimes
(|VV\rangle_{34} + r e^{i\phi}|HH\rangle_{34}).
\end{eqnarray}  
Furthermore, after the  beams pass through the two horizontal polarizing
beam splitters( denoted by PBS1), with perfect synchronization\cite{zuo},
the state is
\begin{eqnarray}
|\chi'\rangle = \frac{1}{1+r^2}(|H\rangle_{1'}|H\rangle_{2'}+ r e^{i\phi}
|V\rangle_{3'}|V\rangle_{4'})\otimes
(|V\rangle_1|V\rangle_2+ r e^{i\phi}
|H\rangle_3|H\rangle_4).
\end{eqnarray}  
This can be recast to the summation of three orthogonal terms: 
\begin{eqnarray}\label{chi}
|\chi'\rangle = \frac{1}{1+r^2}(|A\rangle + |B\rangle + |C\rangle)
\end{eqnarray}
where
\begin{eqnarray}
|A\rangle = r e^{i\phi}( |H\rangle_{1'}|H\rangle_{2'}|H\rangle_{3'}|H\rangle_{4'}
+|V\rangle_{1'}|V\rangle_{2'}|V\rangle_{3'}|V\rangle_{4'} );
\end{eqnarray}
\begin{eqnarray}
|B\rangle = |H\rangle_{1'}|V\rangle_{1'}|H\rangle_{2'}|V\rangle_{2'}; \\
|C\rangle= r^2 e^{2i\phi} |H\rangle_{3'}|V\rangle_{3'}|H\rangle_{4'}|V\rangle_{4'}.
\end{eqnarray}
Obviously, term $|B\rangle$ means that there is no photon in beam 4' therefore
this term will never click detector $D_w$. Consequently term $|B\rangle$ will
never cause the two fold coincidence. Similarly, term $|C\rangle$  means there
is no photon in beam  1' therefore it will never cause the required
two fold coincidence either. Since neither term $|B\rangle$ nor term
$|C\rangle$ will cause the required two fold coincidence, we disregard these 
two terms and only consider term $|A\rangle$ hereafter.
The overall factor $\frac{1}{1+r^2}$ or $r e^{i\phi}$ plays no role in  any
physical results therefore is replaced by a normalization factor hereafter.

The HWP2 takes a $\pi/4$ rotation of the beam's polarization, i.e.,
it changes $|H\rangle$ into $\frac{1}{\sqrt 2}(|H\rangle +|V\rangle)$
and $|V\rangle$ into $\frac{1}{\sqrt 2}(|H\rangle -|V\rangle)$.
After the beams pass through HWP2,  state $|A\rangle$ is evolved to
\begin{eqnarray}
 |\Phi^+\rangle_{1''4''} |\Phi^+\rangle_{2'3'}+
 |\Psi^+\rangle_{1''4''} |\Phi^-\rangle_{2'3'}
\end{eqnarray}
where $|\Phi^{\pm}\rangle_{ij}
=\frac{1}{\sqrt 2}( |H\rangle_i|H\rangle_j\pm |V\rangle_i|V\rangle_j$ and
 $|\Psi^{\pm}\rangle_{ij}
=\frac{1}{\sqrt 2}( |H\rangle_i|V\rangle_j\pm |V\rangle_i|H\rangle_j$. 
After pass through the two PBS2 in the figure,  $|A\rangle$  
is evolved to  state
\begin{eqnarray}\label{xyzw}
|x\rangle|w\rangle |\Phi^+\rangle_{2'3'} + |y\rangle|z\rangle 
|\Phi^+\rangle_{2'3'} +
(|x\rangle|z\rangle +|y\rangle|w\rangle)|\Phi^-\rangle_{2'3'}  
\end{eqnarray}
where $|s\rangle$ denote the state of one photon in beam $s$, $s$ can be
$x,y,z$ or $w$.  From the above formula we can see that only the first term will cause the two fold coincidence event that both detectors are clicked. And we see that this term indicates a maximally event-ready entangle state between
beam 2' and 3', i.e. $\frac{1}{\sqrt 2}(|H\rangle_{2'}|H\rangle_{3'}+
|V\rangle_{2'}|V\rangle_{3'})$. Note that our result here is independent of the parameters $r,\phi$.
Also, the quality of the outcome is not affected by the efficiency of the 
photon detectors.  Actually, the overall
efficiency of the scheme can be increased by 4 times with two more
photon detectors $D_y, D_z$, detecting the beam $y$ and beam $z$ respectively.

To really produce the event-ready entanglement through our
purification scheme  one need the deterministic
supply of the requested raw states. 
This  is rather challenging a task.
However, even without such a deterministic supply, one can still
experimentally demonstrate that our scheme $can$ produce the event-ready
maximally entangled pair $if$ the deterministic supply is offered. 
In a real experiment to demonstrate our scheme, 
one can probabilistically produce the requested non-maxmally entangled
initial state by  the type II 
SPDC process or other processes
for the pair 1,2 and the pair 3,4\cite{ypra,para1,para}. 
In such a case it is also possible
that actually the pair 1,2 contains nothing while 3,4 contains two pairs,
and vice versa.
However, one may  discard those cases by a post selection, i.e., 
by studying  the 4 fold coincidence. We must point out that, this post selection method does not affect the non-post-selection  nature of our scheme:
Limitted to the imperfection in the raw state preparation,
 the  experimental motivation here 
is not the very ambitious one to $really$ produce the 
event-ready maximally entangled pairs, instead, it is to
verify that this set-up $can$ produce the event-ready maximally entanglement 
pairs {\it provided that} the requested raw state $\int |r,\phi\rangle\langle r,\phi|dr d\phi $ is supplied deterministically. 
The post selection here is only to exclude those events where a wrong raw 
state had been produced, it is not used  
to exclude the corrupted outcome due to
the imperfection of the devices, such as the yes-no photon detector. 
Here the manufacturer can claim safely to their customers that the set-up
produces event-ready EPR pairs provided that the customers input the requested
raw states.

With the SPDC process, the 
 initial state prepared for pair 1,2 and pair 3,4 by Fig.1,2 is
\begin{eqnarray}\label{initial}
|in\rangle = 2|raw_1\rangle + \sqrt 3 |u_1\rangle + \sqrt 3 |u_2\rangle
\end{eqnarray}
where state $|raw_1 \rangle= \frac{1}{1+r^2}(a_{1H}^\dagger a_{2H}^\dagger+
re^{i\phi}a_{1V}^\dagger a_{2V}^\dagger )(a_{3H}^\dagger a_{4H}^\dagger+
re^{i\phi}a_{3V}^\dagger a_{4V}^\dagger)|0\rangle$ is just the requested
raw state of our entanglement concentrator,
$|u_1\rangle=\frac{1}{(1+r^2)\sqrt 3}(a_{1H}^\dagger a_{2H}^\dagger+
e^{i\phi}a_{1V}^\dagger a_{2V}^\dagger )^2|0\rangle$ and
$|u_2\rangle=\frac{1}{(1+r^2)\sqrt 3}(a_{3H}^\dagger a_{4H}^\dagger+
e^{i\phi}a_{3V}^\dagger a_{4V}^\dagger)^2|0\rangle$.
 Note that we have omitted
the overall normalization factor which plays no role here.
 
Consider the scheme in Fig.2.  Note that R there is a phase shifter which
offers a shift $\theta$ randomly chosen from $\{\theta_1=0, \theta_2=\pi/2,
\theta_3= -\pi/2, \theta_4=\pi\}$. Therefore  the 
 input state is now
\begin{eqnarray}
\rho = \frac{1}{4}\sum_{j=1}^4 |\theta_j\rangle\langle \theta_j|
\end{eqnarray}
and $|\theta_j\rangle = 
2 e^{i\theta_j}|raw_1\rangle + \sqrt 3 e^{2i\theta_j}|u_1\rangle + \sqrt 3 |u_2\rangle $.
  By a straightforward calculation we have
\begin{eqnarray}\label{rho}
\rho= 4 |raw_1\rangle\langle raw_1| + 3 |u_1\rangle\langle u_1| + 3 |u_2\rangle\langle u_2|.
\end{eqnarray}
Therefore the input state is now in a classical probabilistic mixture of
the requested raw state $|raw_1\rangle$ and the unwanted states $|u_1\rangle$,
 $|u_2\rangle$. This is to say, if we input the state for $1000$ times, $400$ of
them are the requested raw state $|raw_1\rangle$ and 300 of them are
$|u_1\rangle$ and 300 of them are $|u_2\rangle$. 
 We shall first observe the consequence $C_1$ caused by 
only inputing
state $|u_1\rangle\langle u_1|$ for 300 times,
 then observe the consequence $C_2$ caused
by only inputing state $|u_2\rangle\langle u_2|$ for 300 times. We finally
 observe the consequence
caused by sending $\rho$ for 1000 times. This consequence, with subtraction of 
$C_1$ and $C_2$, is the net consequence caused by the 400 times inputs
of requested raw state
$|raw_1\rangle\langle raw_1|$. \cite{remark1}
 
To only input state $|u_1\rangle\langle u_1|$ 
we block beam 3,4 and input only the pair 1,2  into the set-up in Fig.2.
Suppose we input 300 copies of state  $|u_1\rangle\langle u_1|$. 
In such a case, the  only type of 4-fold clicking is $(D_x, D_w, D_{2H}, D_{3V})$.
Suppose we have observed $N$ times of such kind 4-fold clicking.
Other types of four fold simultaneous clicking,
$(D_x, D_w, D_{2H},D_{3H})$ or $(D_x, D_w, D_{2V}, D_{3V})$  
or $(D_x, D_w, D_{2V}, D_{3H})$ will never be observed\cite{remark8}. 
With the same source, if we
rotate both beam 2' and beam 3' by $\pi/4$ before 
detection, we shall find  
$N/4$ times of  4-fold events for each of the simultaneous clicking of  
$(D_x, D_w, D_{2H}, D_{3V})$,
$(D_x, D_w, D_{2V}, D_{3H})$,$(D_x, D_w, D_{2H}, D_{3H})$ and 
$(D_x, D_w, D_{VH}, D_{3V})$ .

Similarly, if we block beam 1 and 2 and input state $|u_2\rangle\langle u_2|$ for 300 times,
we shall observe $N$ times of 4-fold clicking $(D_x,D_w,D_{2V},D_{3H})$
and no other types of 4-fold clicking.
And
also, if we insert a $\pi/4$ HWP to both beam 2' and beam 3'
and repeat the test, we shall find $N/4$ times of 
4-fold events for each of the simultaneous clicking of  
$(D_x, D_w, D_{2H}, D_{3V})$,
$(D_x, D_w, D_{2V}, D_{3H})$,$(D_x, D_w, D_{2H}, D_{3H})$ and 
$(D_x, D_w, D_{VH}, D_{3V})$ .

Keeping these facts in the mind we now consider the test by Fig. 2
 with
the input state $\rho$ given 
by Eq.(\ref{rho}). We input state $\rho$ for 1000 times. {\it Physically, this is equivalently to
input each of state $|raw_1\rangle$, $|u_1\rangle$ and $|u_2\rangle$ 
 for 400 times, 300 times and 300 times, respectively.}
We shall observe $N$ times four fold coincidence events
events of each of $(D_x, D_w, D_{2H}, D_{3V})$, 
$(D_x, D_w, D_{2V}, D_{3H})$,  $(D_x, D_w, D_{2H}, D_{3H})$,  
$(D_x, D_w, D_{2V}, D_{3V})$ without inserting HWP to beam 2' or beam 3'. 
However, as we have already known that, among all these 4-fold events, the 300
inputs of $|u_1\rangle\langle u_1|$ have caused $N$ times of 4-fold
clicking $(D_x, D_w, D_{2H}, D_{3V})$; the 300 inputs of $|u_2\rangle\langle u_2|$
have caused  $N$ times  4-fold
clicking of $(D_x,D_w,D_{2V},D_{3H})$. This is to say, the only 4-fold events caused
by the 400 times inputs of $|raw_1\rangle\langle raw_1|$ are just the    
$(D_x, D_w, D_{2H}, D_{3H})$ and  
$(D_x, D_w, D_{2V}, D_{3V})$. This shows the following conclusion:
\\{\it Conclusion 1:}  We shall only observe 4-fold events of
either  $(D_x, D_w, D_{2H}, D_{3H})$ or
$(D_x, D_w, D_{2V}, D_{3V})$, $if$ the input is deterministically 
$|raw_1\rangle\langle raw_1|$. 

We then insert a $\pi/4$ HWP to both
 beam 2' and beam 3'. 
This time we shall find $N/2$ four
fold events of  $(D_x, D_w, D_{2H}, D_{3V})$, 
 $N/2$ four
fold events of $(D_x, D_w, D_{2V}, D_{3H})$, $3N/2$ times of  four
fold events of  $(D_x, D_w, D_{2H}, D_{3H})$ and  $3N/2$ times of  four
fold events of  $(D_x, D_w, D_{2V}, D_{3V})$. However, as we have already 
known from the previous paragraphs,  
in such a case the 300 times of inputs of  $|u_1\rangle\langle u_1|$ and
$|u_2\rangle\langle u_2|$ have caused
$N/2$ times of four fold coincidence events for each of the  four types
of four fold simultaneous clicking. Therefore the net result
caused by the 400 time inputs of $|raw_1\rangle\langle raw_1|$ is only the four fold events of
 $(D_x, D_w, D_{2H}, D_{3H})$ and  $(D_x, D_w, D_{2V}, D_{3V})$, each appearing $N$ times.
This  shows the following conclusion:
\\{\it Conclusion 2:} After inserting a  $\pi/4$ HWP to both
 beam 2' and beam 3', one shall only  only observe 4-fold events of
either   $(D_x, D_w, D_{2H}, D_{3H})$ or  $(D_x, D_w, D_{2V}, D_{3V})$,
$if$ the input is 
$|raw_1\rangle\langle raw_1|$. 

Combining {\it conclusion 1} and {\it conclusion 2}  we conclude that no matter whether we 
 insert $\pi/4$ HWP to beam 2', 3',
the net four fold
events caused by   $|raw_1\rangle\langle raw_1|$ 
 are only $(D_x, D_w, D_{2H}, D_{3H})$ and  $(D_x, D_w, D_{2V}, D_{3V})$.
Therefore the non-post-selection entanglement concentrator is demonstrated.
In carrying out the experiment, one should
make sure $N$ large enough so 
that $\frac{\sqrt N}{N}<<1$,  to reduce
the statistical error. This requires that the set-up
has to be stable for several hours\cite{ustc}.   

{\bf Acknowledgement:} We thank Prof. H. Imai  for support. 
We thank Dr K. Matsumoto and Dr A. Tomita for discussions. 

\begin{figure}
\begin{center}
\epsffile{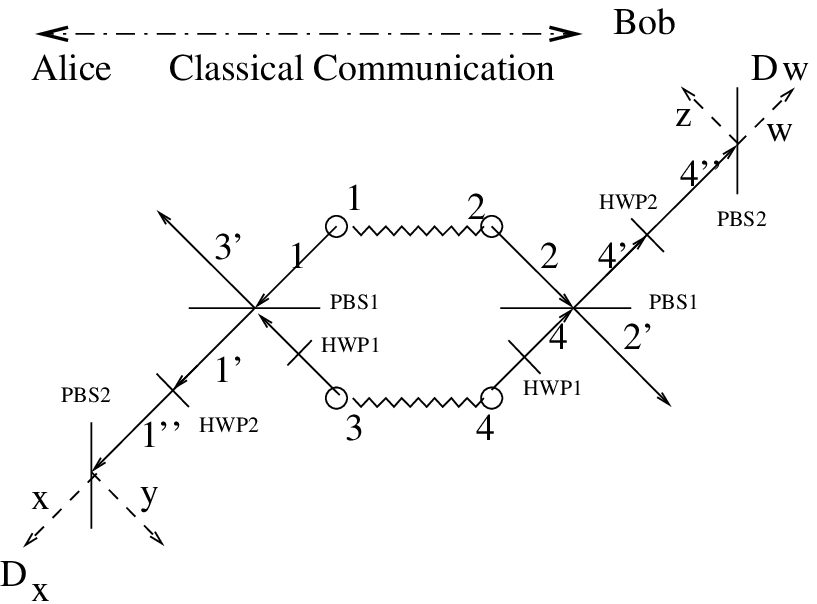}
\end{center}
\caption{Non-post-selection
quantum entanglement concentration by practically existing
devices of  linear optics. The two fold coincidence event of both
detector $D_x$ and detector $D_w$ being clicked indicates that a maximally entangled
state is produced on beam 2' and 3'.
PBS: polarizing 
beam-splitter.  Here
HWP1 rotates the polarization by $\pi/2$, HWP2 rotates the polarization
by $\pi/4$. }
\end{figure}
\begin{figure}
\begin{center}
\epsffile{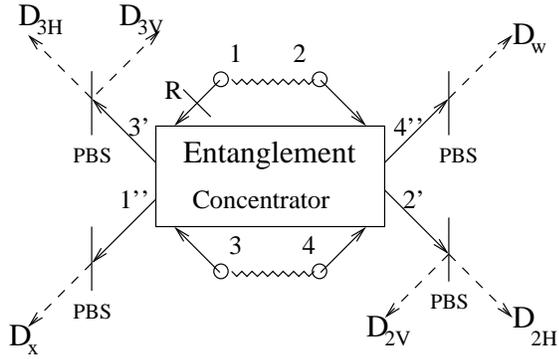}
\end{center}
\caption{A schematic al scheme to verify that all the four fold
coincidence events caused by the 
state $|raw_1\rangle\langle raw_1|$ are
$D_x, D_w, D_{2H}, D_{3H}$ or $D_x, D_w, D_{2V}, D_{3V}$, no matter
whether we take a $\pi/4$ polarization rotation to beam 2' and beam 3' before
the detection. $ R$ is a phas a  phase shift randomly chosen
from $\{0, \pi/2, -\pi/2 ,\pi \}$.
 The square box of ``Entanglement Concentrator'' is just the set-up given by
Fig.1 with the two PBS2 and two photon detectors being cut off.} 
\end{figure}

\begin{thebibliography}{99}
\bibitem{remark1}Consider an analog classical story: 
The fake coin and the good coin cannot be distinguished by eyes. Suppose we have produced a machine and we 
want to demonstrate the claim that whenever you insert a good coin, it produces a bottle of wine.
However, the machine itself has a property that if you insert a fake coin, it produces a bottle of water.
We don't have any supply of pure good coins. But we have pure supply of fake coins, supply $B$ and
mixed supply of half good coins and half fake coins, supply $A$. We input 200 coins from supply $A$ and find 100 bottles of 
wine and 100 bottles of water. We then input 100 coins from supply $B$ and find 100 bottles of water. Our claim is
verified by the combination of these two facts. In our main text the role of
$\rho$ is similar to  supply $A$ while $|u_{1,2}\rangle\langle u_{1,2}|$ is similar to supply $B$.
\bibitem{remark8}
With beam 3 and beam 4 being blocked, the
 input state is now $|u_1\rangle$ and the  state
of beam 1'',2',3' and 4'' 
is
$
\frac{1}{\sqrt 3}[|2H\rangle_{1''}|2H_{2'}\rangle+ r^2e^{2i\phi}
|2V\rangle_{3'}|2V\rangle_{4''}
+\frac{1}{2}re^{i\phi}(|H\rangle_{1''}+|V\rangle_{1''})|H\rangle_{2'}
|V\rangle_{3'}(|H\rangle_{4'}-|V\rangle_{4''})]
$
the four fold event is uniquely caused by the term
$\frac{1}{2}re^{i\phi}(|H\rangle_{1''}+|V\rangle_{1''})|H\rangle_{2'}
|V\rangle_{3'}(|H\rangle_{4'}-|V\rangle_{4''})$.
Whenever we see a four fold event here, it must be   
$(D_x, D_w, D_{2H}, D_{3V})$.
\end{thebibliography}
\end{document}